\documentclass[letter,scriptaddress,twocolumn, prb,showpacs]{revtex4}

	\usepackage{amsmath}
	\usepackage{makeidx}
	\usepackage{soul}
	\usepackage{array}
	\usepackage{booktabs}
	\usepackage{amsfonts}
	\usepackage[ansinew]{inputenc}
	\usepackage[usenames,dvipsnames]{pstricks}
	\usepackage{subfigure}
	\usepackage{epsfig}
	\usepackage{pst-grad}
	\usepackage{pst-plot}
    \usepackage[normalem]{ulem}
    \usepackage{multibbl}
	\usepackage[colorlinks,hyperindex]{hyperref}
	\usepackage{lipsum}
	\hypersetup
	{
		colorlinks,
		citecolor=black,
		linkcolor=black,
		urlcolor=black,
	}

\newcounter{mybibstartvalue}
\makeindex

\begin{document}

\title{Motion and teleportation of polar bubbles in ultra-thin ferroelectrics}

\author{S. Prokhorenko$^1$}
\email{prokhorenko.s@gmail.com}
\author{Y. Nahas$^1$}
\author{Q. Zhang$^2$}
\author{V. Govinden$^2$}
\author{N. Valanoor$^2$}
\author{L. Bellaiche$^1$}
\affiliation{$^1$Physics Department and Institute for Nanoscience and Engineering, University of Arkansas, Fayetteville, Arkansas 72701, USA}
\affiliation{$^2$School of Materials Science and Engineering,
The University of New South Wales,
Sydney, New South Wales 2052, Australia}

\date{\today}

\maketitle
\textbf{Polar bubble domains are complex topological defects akin to magnetic skyrmions~\cite{Rossler2006} that can spontaneously form in ferroelectric thin films~\cite{Igor,Lai,Zhang2017} and superlattices~\cite{Ramesh2019}. They can be deterministically written and deleted~\cite{Zhang2019} and exhibit a set of properties, such as sub-10 nm radius and room-temperature stability, that are highly attractive for dense data storage and reconfigurable nano-electronics technologies.
However, possibilities of \textit{controlled motion} of electric bubble skyrmions, a critical technology requirement~\cite{Ramesh2019} currently remains missing. Here we present atomistic simulations that demonstrate how external electric-field perturbations can induce two types of motion of bubble skyrmions in low-dimensional tetragonal PbZr$_{0.4}$Ti$_{0.6}$O$_3$ systems under residual depolarizing field. Specifically, we show that, depending on the spatial profile and magnitude of the external field, bubble skyrmions can exhibit either a continuous motion driven by the external electric field gradient or a discontinuous, teleportation-like, skyrmion domain transfer. These findings provide the first analysis of dynamics and controlled motion of polar skyrmions that are essential for functionalization of these particle-like domain structures.
}


\begin{figure}[ht!]
\centering
\includegraphics[scale=0.21]{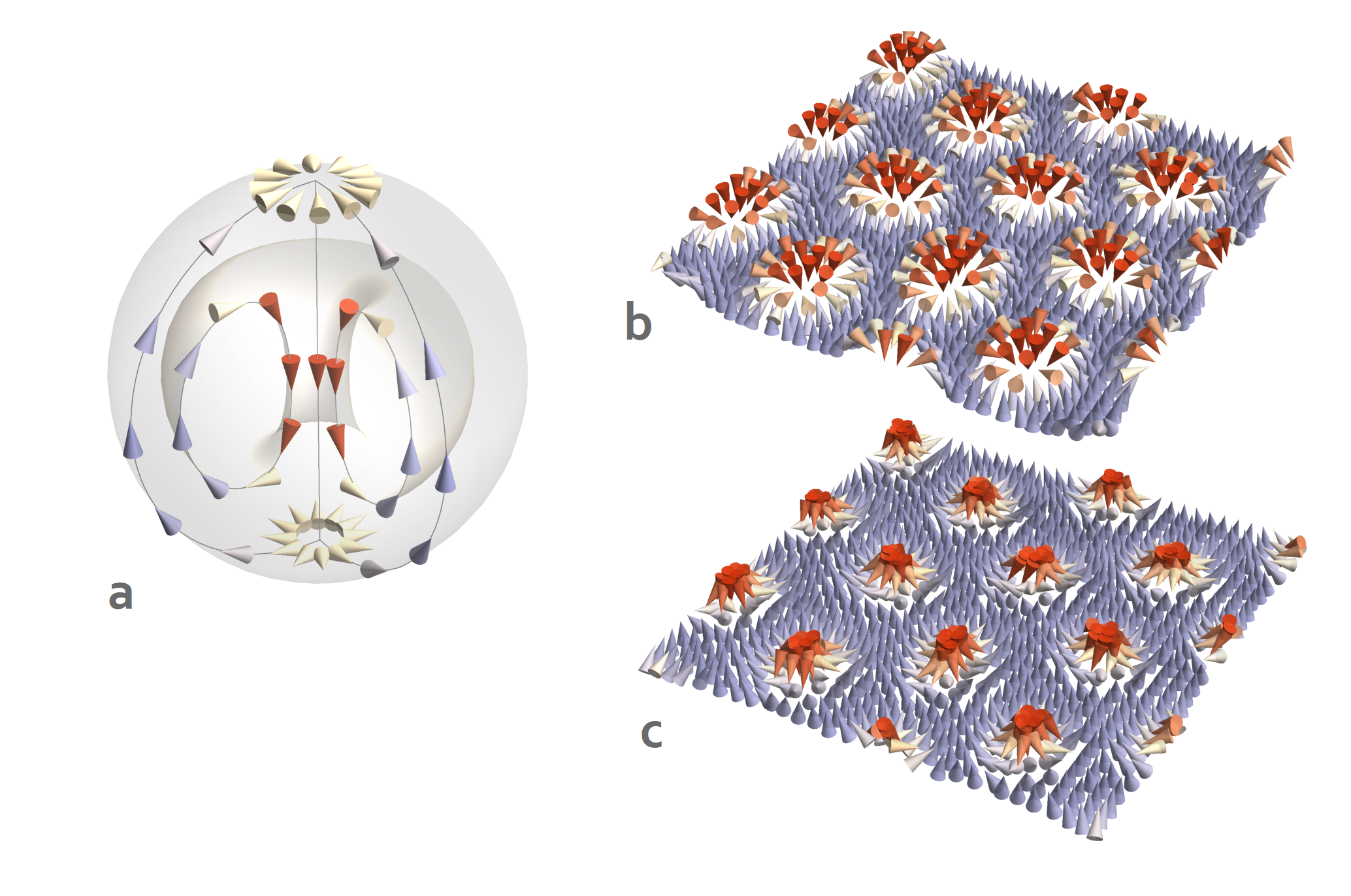}
\caption{\textbf{Dipolar structure of bubble domains.} (a) Schematic representation of the low-temperature dipolar structure of the bubble domains. The dipolar pattern possesses a revolution symmetry axis passing through the ``north'' and ``south'' poles of the spherical domain boundary. This symmetry axis also corresponds to the locus of dipoles oriented parallel to the $z$ axis. At the domain boundary, electric dipoles are tangential to the domain surface meridians and, in accordance with the hedgehog theorem~\cite{algTopology}, form two-dimensional monopole patterns in the close vicinity of the poles. The ``meridional'' cross-section reveals an inner tubular vortex structure with dipoles rotating from $+z$ to $-z$ direction within the plane of the cut. Panels (b) and (c) show the dipolar structure of the bubble domain lattice within $z=2$ and $z=4$ planes, respectively, obtained from $32\times 32\times 5$ supercell simulations of  Pb(Ti$_{0.4}$,Zr$_{0.6}$)O$_3$ slab at 10K subjected to a  $-2\%$ compressive misfit strain and an external electric field $E_z$ of $41\times 10^7$ V/m. The cross-sections of bubble domains (dipoles highlited in red and orange) can be readily identified as polar skyrmions~\cite{Ramesh2019}, although with an opposite sense of rotation within the planes below ($z=2$) and above ($z=4$) the equatorial plane (see panel (a)).
}
\label{fig:bubblestruct}
\end{figure}

Owing to their nanometer length-scale, stability and electric field sensitivity, dipolar topological patterns in ferroelectrics hold an extraordinary technological potential for information storage and processing that challenges the existing spintronics and skyrmionics applications. Of particular interest are the ferroelectric bubble domains, theoretically predicted~\cite{Igor,Lai} more than a decade ago to spontaneously form in Pb(Zr,Ti)O$_3$ (PZT) thin films and recently observed at room-temperature in Pb(Zr,Ti)O$_3$/SrTiO$_3$ sandwich structures~\cite{Zhang2017,Zhang2019} and PbTiO$_3$/SrTiO$_3$ superlattices~\cite{Ramesh2019}. Such domains were also latterly shown~\cite{Ramesh2019} to carry an integer skyrmionic charge that, in addition to the axial symmetry of these nano-patterns, triggered a fresh recognition of polar bubbles as the electric counterparts of magnetic skyrmions. This topological insight has further raised the question~\cite{Ramesh2019} of whether one can envision a bubble-based analogue of the magnetic racetrack memory~\cite{racetrack} that would allow for electric-field bit manipulation and superior data stability. However, in order to leverage electric skyrmions in such novel device architectures, it is essential to understand several, yet unknown, properties of bubble domains states. For instance, many functionalization routes, including the electric racetrack concept, would rely not only on the possibilities of moving bubble domains but also on the ability of controlling the density of bubble skyrmions thus far predicted and observed to form either a domain lattice pattern~\cite{Lai,Igor} or dense domain ensembles~\cite{Zhang2017,Ramesh2019}. Here, we address both of these currently unexplored problems via large-scale atomistic simulations of bubble domains in two-dimensional Pb(Zr$_{0.4}$,Ti$_{0.6}$)O$_{3}$  structures under residual depolarizing field. 

Bubble domains in tetragonal PZT thin films have been shown to possess an intricate dipolar structure~\cite{Igor,Lai} (see Fig. 1a) that is characterized by a tangential orientation of the dipoles lying on the outer spherical boundary of the domain and a non-zero domain polarization stemming from collinear dipoles located at the polar axis of the bubble. The polar $z$ axis, as well as the overall bubble polarization, are both found to be oriented perpendicular to the free surface of the ferroelectric structure. Furthermore, the latitudinal cross-sections of the bubble reveal the continuity of dipolar rotations upon moving away from the bubble center, with the core out-of-plane polarization gradually re-orienting towards the anti-parallel polarization of the embedding matrix. Such rotations in different $z$ planes create nested toroidal surfaces tangent to the dipolar vector field within the bubble (see Fig.~1a). This inner dipolar structure resembles the geometry of electric field lines of a standalone electric dipole.
%
Similarly to the case of PbTiO$_3$/SrTiO$_3$ superlattices~\cite{Ramesh2019}, our calculations of the Pontryagin winding number~\cite{Yousra2015}  confirm the topological nature of the bubble domains in 5 u.c. thick Pb(Zr$_{0.4}$Ti$_{0.6}$)O$_3$  slabs under electric boundary conditions mimicking realistic electrodes~\cite{DepField} (80\% screening of surface bound charges) and $-2.65\%$ compressive strain. Specifically, for each of the $z$ planes of the bubble we found a +1 integer skyrmionic charge, hence confirming the non-trivial topology of the corresponding two-dimensional dipolar textures. This result shows that the bubble domains in PZT thin layers are thus topologically equivalent to polar skyrmions reported in superlattice structures~\cite{Ramesh2019}.
%
%
Interestingly, our calculations of the winding number also show that the sign of the integer skyrmion charge does not change from one latitude to another while the sense of dipolar rotation does reverse at the equatorial plane as can be seen from simulation results presented in Figs. 1(b) and 1(c). This geometric particularity constitutes one of the differences that distinguish bubble skyrmions from their magnetic siblings - two-dimensional skyrmions and quasi two-dimensional skyrmionic tubes~\cite{SkyrmionsBook}. Another key dissimilarity which remains unappreciated thus far, is the decrease of the $z-$cross section radius upon moving away from the bubble equator which results in oppositely charged two-dimensional monopoles located at the ``north'' and ``south'' poles of the domain boundary (see Fig.~1a). 
This pair of surface defects is reminiscent of boojums~\cite{boojums1,boojums3} and renders the topological structure of the electrical skyrmion bipolar.  It concentrates the electric bound charge rising from non-zero divergence~\cite{LandauED} of the electric dipole moment, which further highlights the overall ``super-dipole'' character of bubble domains in PZT systems.

\begin{figure}[ht!]
\centering
\includegraphics[scale=0.26]{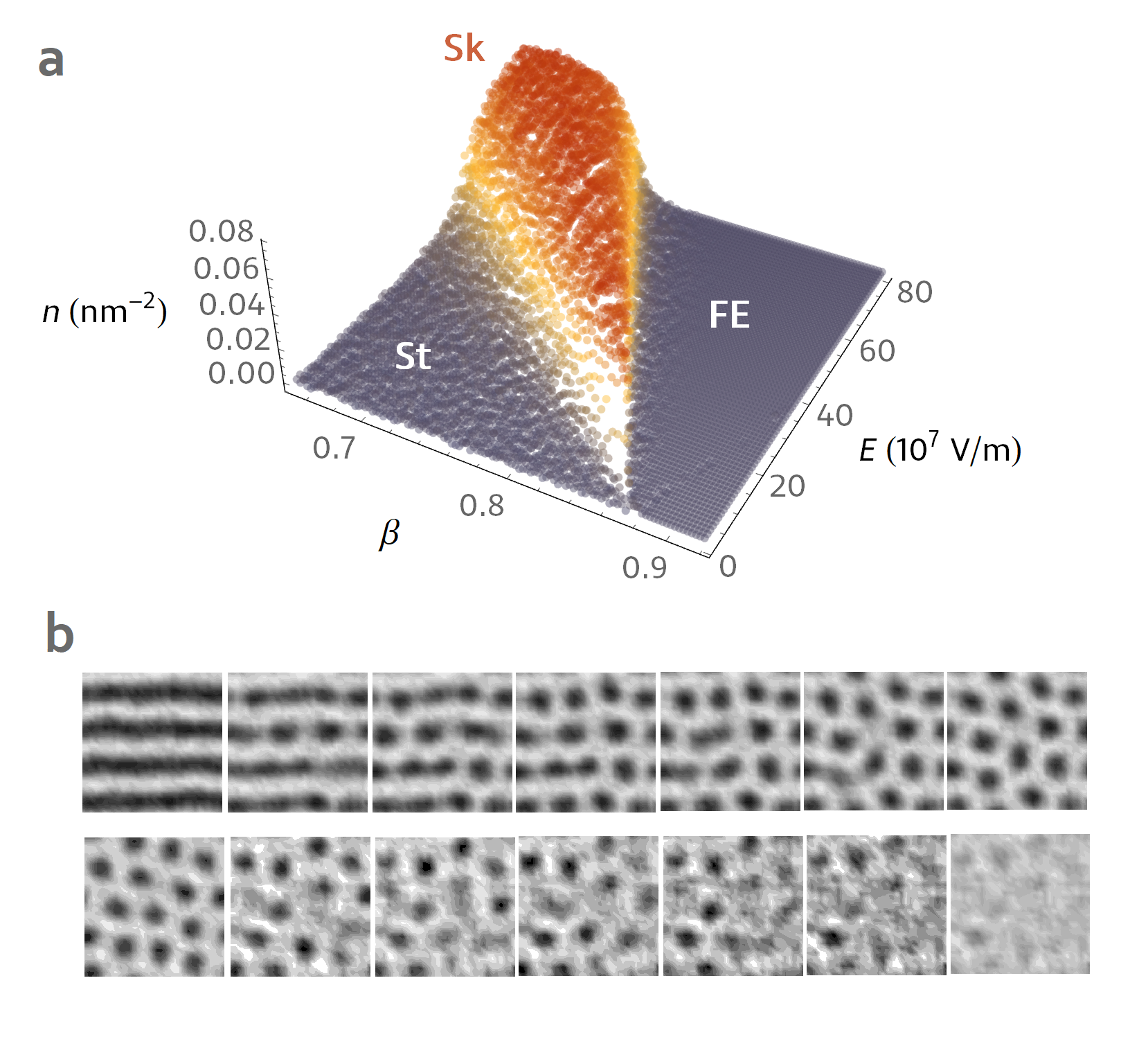}
\caption{\textbf{Electric field-screening phase diagram.} (a) Variation of the density $n$  of skyrmion bubble domains depending on the external electric field magnitude and the effective screening $\beta$. The region of high $n$ values corresponds to the skyrmion lattice phase ``Sk'' which delimits the vortex tube or stripe phase ``St'' (low $\beta$ and electric field values) from the polar ferroelectric state ``FE'' (high $\beta$ and high electric field magnitudes). For both ``St'' and ``FE'' phases the bubble domain density $n$ is zero. Panels (b) and (c) show the evolution of the domain pattern in the vicinity of St-Sk and Sk-FE transitions, respectively, and demonstrate that the control of density of bubble domains can be achieved by application of external bias field.}
\label{fig:phdiag}
\end{figure}

Once the focus is shifted from the microscopic dipolar structure to the global properties of the system, we notice that, in contrast to the stripe~\cite{Lai,Yadav2016} (vortex tube) domain patterns, the bubble domain state is macroscopically polar. The corresponding polarization vector is aligned with the polarization of the embedding matrix since the volume occupied by bubble domains is lower than that occupied by the matrix domain. This simple fact suggests that in order to stabilize bubble domains the system needs to break the $z$-axis inversion symmetry imposed by strong depolarizing fields which favor zero net polarization. The symmetry can be trivially lowered by applying an external electric field perpendicular to the film's surface~\cite{Lai}, which was shown to indeed transform the stripe domains into polar bubbles at 10K in PZT thin films. Such effect is remarkably similar to external magnetic field stabilization~\cite{SkyrmionsBook} of magnetic skyrmions in thin films of chiral magnets. Another possibility resides in triggering the spontaneous symmetry breaking. This second option can be achieved by reducing~\cite{Igor} the depolarizing field strength through the introduction of buffer layers that allow to weaken the discontinuity of polarization. The stabilization of bubble domains via either of these mechanisms has been previously numerically demonstrated  at low temperatures~\cite{Lai,Igor}. However, realistic experimental conditions often involve both factors, as in the case of the bubble domains observed~\cite{Zhang2017} in PZT bilayers buffered with SrTiO$_3$, where the intrinsic asymmetry of the grown sandwich structure gives rise to built-in electric field acting as an external bias~\cite{Zhang2019}. In order to better understand the symmetry breaking occurring under a more realistic, combined action of these factors, we have numerically obtained the screening-electric field phase diagram of compressively strained Pb(Zr$_{0.4}$Ti$_{0.6}$)O$_3$ layers at room temperature. For this, we have used Metropolis Monte Carlo annealing simulations with a first-principles-based effective Hamiltonian model~\cite{Igor} that has been proven in the past to be accurate in predicting finite-temperature structure of ultrathin PZT films. A supercell consisting of 32$\times$32$\times$5 unit cells with in-plane periodic boundary conditions was used to mimick a PZT film geometry with 2 nm thickness. An effective parameter $\beta$ is introduced~\cite{DepField} to simulate partial screening boundary conditions along the $z$ axis. The screening efficiency $\beta$ corresponds to the fraction of the bound charge screened at the surfaces of the two-dimensional structure with $\beta=0$ corresponding to ideal open circuit and $\beta=1$ to ideal short circuit boundary conditions. The external field is applied along the $z$ axis perpendicularly to the film surface. The resulting phase diagram displaying the bubble skyrmion density plot is presented in Fig.~2a. As expected~\cite{Igor,Lai}, the low effective screening $\beta$ and low electric field $E$ values grant thermodynamic stability to the stripe domain state with $\sim$2.4 nm in-plane period (leftmost upper panel in Fig.~2b), while high $\beta$ and $E$ values yield a homogeneously polarized sample. Assuming external conditions such that the stripe pattern is an equilibrium state and increasing the external applied field magnitude triggers two consequent phase transitions. The first transition from stripe domains to bubble skyrmion state is achieved by a gradual disconnection of the vortex tubes (dash-dot domain pattern) followed by the re-organization of the linear arrangement of bubble skyrmions into a hexagonal lattice. An example of such evolution under increasing external field magnitude is presented in the upper row of panels in Fig.~2b. 
Further increasing the applied field magnitude eventually destroys the skyrmion lattice and results in a homogeneous structure polarized along the external field. We find that this transition is also gradual - the bubbles are, ``one-by-one'', erased by the external field with corresponding domain lattice vacancies concentration increasing with the magnitude of the bias (lower set of structure snapshots in Fig.~2b). The gradual character of both described transitions therefore allows to practically control the density of bubble skyrmion domains. This can be clearly seen from Fig.~2a where the region of skyrmion lattice stability corresponds to the area of the skyrmion density plot highlighted in red. The left and right slopes of the plot designated by yellow color correspond to the stripe-to-bubble and bubble-to-monodomain transitions, respectively. Tuning the values of external parameters within these regions allows to achieve skyrmion surface densities from 0 to 8 bubbles within a 10 nm$\times$10 nm area of the film. As can be also seen from Fig.~2a, the low- and high-field transitions occur on the straight lines with negative slopes in the $\beta$-$E$ parameter space. Specifically, the critical bias field magnitudes linearly decrease with increasing screening efficiency. Furthermore, the transition lines converge when the screening efficiency $\beta$ is increased. In other words, higher $\beta$ values yield a narrower $E$-field range of stability of the bubble skyrmion lattice until the intersection of the transition lines at $\beta=\beta_c=0.875$ point on the $E=0$ V/m axis. This intersection pinpoints the $\beta$-induced  first-order phase transition from stripe domain structure to the monodomain ferroelectric state stable under boundary conditions sufficiently close to the short-circuit case. Interestingly, at this zero bias field phase transition, the described transformations shown in Fig.~2b also occur sequentially, but in time. Specifically, during the nucleation process at $\beta=\beta_c$ and $E=0$ V/m where the stripe state transforms into a monodomain ferroelectric, the temporal evolution of the system is such that it transits through the same states that are found in stripe-bubble lattice-ferroelectric transition sequence at $\beta<\beta_c$ under increasing $E$.
%
\begin{figure*}[t]
\centering
\includegraphics[scale=0.45]{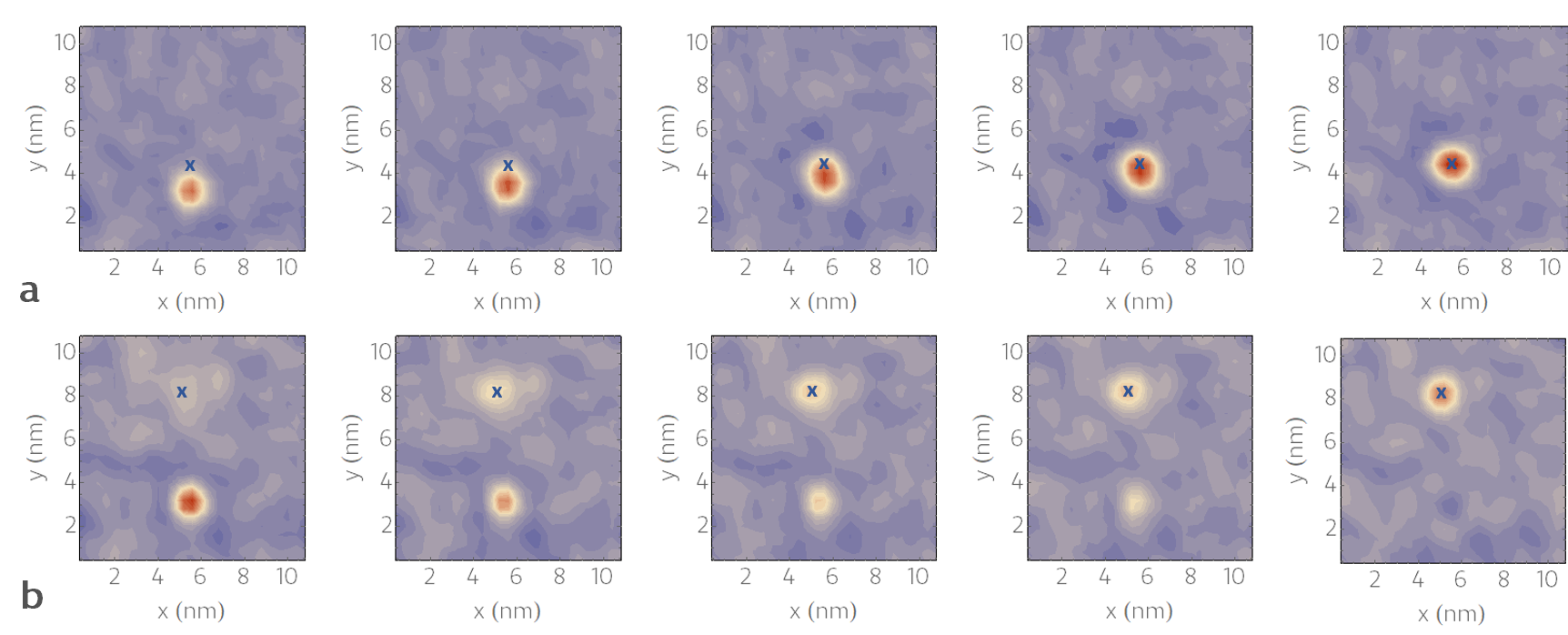}
\caption{\textbf{Electric-field-induced motion of polar bubbles.} Panel (a) shows the simulation of short-range displacement of bubble domain induced by PFM tip located 2 nm away from the bubble center (position of the tip indicated with a blue cross). The driving field profile with a maximum local value of 21 $\times  10^7$ V/m is assumed. As it can be seen, the bubble skyrmion displaces from its initial position (left-most image) towards the point right beneath the PFM tip (right-most image) through a stretch-like motion (middle image). Panel (b) shows the evolution of domain structure when PFM tip (blue cross) is applied 5.5 nm away from the bubble center. Here, the local maximum of the field profile is of 110 $\times 10^7 $ V/m. In this case, the local electric field creates a bubble domain beneath the tip, while simultaneously deleting the bubble domain at its initial position. In panels (a) and (b) the left to right sub-panels show the distribution of the $z$ component of polarization at times $t=0$, 0.6~ps, 0.75~ps, 0.8~ps and 1~ps, respectively. Blue to red colors correspond to negative to positive polarization values. All presented simulations results were performed at 10K and a background bias field $E_z$ of $-61\times 10^7 $ V/m. }
\label{fig:motion}
\end{figure*}

The ability to control the number of bubble domains per unit area allows to create a state of particular practical importance where polar skyrmions are quasi-isolated, i.e. where the distance between domains is significantly larger than the domain radius. \emph{However, equally important for applications is the ability to control the displacements of these topological quasi-particles}. For this, one might be tempted to assume that bubble displacements can be induced by the application of an in-plane oriented external field. However, the super-dipolar character of individual bubble domains (Fig.~1a) immediately casts a doubt on the feasibility of such an approach, which is furthermore confirmed by our simulations. As we find, the application of an in-plane electric field leads to a homogeneous tilting of local dipoles along its direction without any change in positions of bubble skyrmions. On the other hand, the application of a \emph{local} electric field oriented along the dipole moment of the bubble (perpendicular to the film surface and oppositely to the direction of external bias), should allow domain displacements since the energy of the system can then be minimized by the motion of the bubble towards the area where such external field magnitude is maximal. 

To test this idea, we have performed molecular dynamics simulations with $\beta=0.80$ and a local driving field profile approximating that of a Piezoresponse force microscopy (PFM) probe~\cite{PFMefield} (see Methods section). The global electric field bias was chosen so as to ensure a single skyrmion per overall supercell area. 
The results presented in Fig.~3 reveal two distinct effects provoked by the considered electric field perturbation depending on its magnitude and the distance of the tip from the center of the bubble domain. Namely, for a tip located up to 3 nm away from the original polar skyrmion and for a maximum electric field magnitude below the coercive field value, we find that the driving field entails a continuous displacement of the polar skyrmion (see Fig.~3a). During such a displacement, the skyrmion shows plasticity as it first elongates towards the tip position before contracting once the polarization within the area beneath the tip is reversed. Interestingly, we find that during this process the overall Pontryagin's topological charge is conserved in contrast to the overall $z$ component of polarization and the surface bound charge. 
In order to estimate the driving force $\pmb{F}$ exerted on the particle-like domain during this process, we note that the energy difference $\delta H$ between the initial and final states can be approximated as
\begin{equation}
\delta H = - \pmb{\delta r}\cdot\pmb{\nabla} (E_z d),
\label{eq:energy_delta}
\end{equation} 
where $\pmb{\delta r}$ denotes the bubble displacement vector, $\pmb{\nabla}$ stands for the gradient operator while $E_z$ and $d$  are the components of external field and domain super-dipole moment perpendicular to the film surface, respectively. From equation (\ref{eq:energy_delta}), noting that $\delta H~\sim-\pmb{F}\pmb{\delta r}$  one obtains
\begin{equation}
\begin{split}
\pmb{F}&=\pmb{\nabla}(E_z d)=d\pmb{\nabla}E_z+E_z\pmb{\nabla}d\\
&=(d+\alpha E_z)\pmb{\nabla}E_z.
\end{split}
\label{eq:driving_force}
\end{equation}
The final expression is derived under the assumption of a linear dependence of $d$ on the $E_z$ with $\alpha$ denoting the corresponding polarizability of the polar skyrmion. Equation~\ref{eq:driving_force} explicitly demonstrates that the driving force is proportional to the gradient of the external electric field, which allows to design field profiles in accordance with the desired motion of the bubble skyrmion.  

In the case where the distance between the original domain center and the tip is larger than 3 nm, our simulations show a skyrmion creation-annihilation process as seen in Fig.~3b. Under these conditions, a maximum driving field magnitude is required to exceed the coercive field. Specifically, we find that a new bubble domain is progressively created beneath the simulated PFM tip, while the polar skyrmion at the initial position progressively disappears. Interestingly, the \emph{transfer} of the switched area is characterized by a synchronous growth and erasure of the new and original domains respectively, as if the polar skyrmion was being teleported towards the PFM tip. Here, by teleporation, we mean the transfer, without traversing of the physical space, of the state and energy associated with the inhomogeneous dipolar order.
The origin of this teleportation-like transfer can be understood by coming back to the skyrmion density plot shown in Fig.~2a. Although the magnitude of the electric field produced by the PFM tip is comparable to that of the electric bias, its spatial extent is limited to only a few nanometers. Therefore, the PFM tip only weakly alters the supercell average value of the external electric field which is not sufficient to change the final equilibrium skyrmion domain density. Therefore the overall polarization is approximately conserved during the process, while the overall skyrmion charge conservation is violated.

Several important notes are due with respect to the described elementary motions. Firstly, both the continuous and discontinuous displacements are triggered only by field magnitudes exceeding a certain threshold value, which is smaller in the case of continuous displacement. Particularly, we find that the activation field for the gradient driven motion is of 20$\times 10^7$ V/m while teleportation-like jumps are set off by fields exceeding $110\times 10^7$ V/m, i.e. the coercive field. Note that the effective Hamiltonian scheme is known~\cite{BinAFE} to overestimate characteristic field magnitudes by a factor of $\sim 20$. Secondly, the continuous displacement example is rather short-distanced and mainly serves as a demonstration of the possibility of gradient driven motion. Longer skyrmion trajectories that are desired for practical applications can be achieved by modifying the driving potential profile  guided by Eq.(2).
In contrast, the skyrmion teleportative motion does not require any additional design of the driving force. However, we would like to note that the PFM tips are known to alter the surface electro-chemistry~\cite{Kalinin} which, in turn, will locally affect the effective screening $\beta$. This could possibly alter the conservation of polarization within the region of interest. Finally, the distance $l$ over which the skyrmion is teleported is limited by domain density $n$, but also by the velocity $v$ of the information propagation in the material. Indeed, by default, the distance between the original and newly nucleated bubble domains cannot exceed the half of the maximum skyrmion separation proportional to $n^{-1/3}$. Furthermore, the time $t_{\uparrow\downarrow}$ needed to reverse the polarization beneath the tip and hence create a bubble domain should be  larger than $l/v$, where $v\sim\xi_0/\tau_0$ is defined by the characteristic correlation length $\xi_0$ of dipolar fluctuations and the relaxation time $\tau_0$. In other words, since the teleportation essentially corresponds to a simultaneous domain deletion and creation, the polarization transfer has to be instantaneous on the time-scale of the reversal process itself, thereby yielding the condition $l\ll \xi_0\tau_{\uparrow\downarrow}/\tau_0$.     

In summary, we report the motion of polar skyrmions in two-dimensional ferroelectrics controlled by an external electric field. This new functional property is realized in the dilute skyrmion density limit that, as we show, can be achieved by tuning either the external electric field bias or interfacial screening conditions. Depending on the magnitude and spatial profile of the driving field, we predict the duality of motion patterns - a continuous skyrmion displacement driven by an electric field gradient and a discontinuous, teleportation-like, jump wherein the original skyrmion domain gradually transfers to a location several diameters away from its initial position without any change of the interstice dipolar structure. We further show that the continuous motion complies with topological charge conservation while the discontinuous one is rather associated with the conservation of polarization. The revealed possibilities to control the density and to move polar skyrmions establish a milestone in the functionalization of polar skyrmions which can lead to novel and highly desired technologies such as electric racetrack memory.

\subsection*{Acknowledgements}
S.P., Y.N., L.B., V. G. and N.V. thank the financial support of the DARPA
Grant No. HR0011727183-D18AP00010 (TEE Program). S.P. and L.B also thank the DARPA Grant HR0011-15-2-0038 (MATRIX program).  Y.N. and L.B.
also acknowledge support of the ARO grant W911NF-16-1-0227.
The research at the University of New South Wales (UNSW) was
supported by an Australian Research Council (ARC) Discovery Project
and partially supported by the Australian Research Council Centre
of Excellence in Future Low-Energy Electronics Technologies (project
number CE170100039) and funded by the Australian Government.

\subsection*{Author contributions}
S.P. performed the simulations. S.P. and Y.N. analysed the results. L. B. initiated and supervised the study. S.P. wrote the first draft of the paper. All authors provided feedback, participated in discussions and editing of the manuscript.

\subsection*{Competing interests} 
Authors declare no competing interests.

\subsection*{Methods}

All simulations are performed for a Pb(Zr$_{0.4}$Ti$_{0.6}$)O$_3$ system  with an effective Hamiltonian model described in Refs~\cite{Lai,Igor,LB1}. The (001) oriented  thin-film or slab geometry of $\sim$2 nm (5 u.c.) thickness is mimicked by a 32$\times$32$\times$5 supercell with periodic boundary conditions imposed along [100] and [010] pseudo-cubic axes. For the results of Fig.~\ref{fig:motion}, the electric boundary conditions along the $z-$axis mimic realistic electrodes that effectively screen 80$\%$ of the polarization-induced surface charges. The depolarizing field in each unit cell is computed using an accurate atomistic model~\cite{DepField} that accurately takes into account inhomogeneities of the polarization gradient distribution and hence accounts for intrinsic size effects in low-dimensional ferroelectrics. For all simulations we assume a compressive strain of - 2.65 $\%$. Such value  approximately accounts for the mismatch of lattice constants of the cubic phases of strontium titanate (STO) and PZT. 
A first-principles-based effective Hamiltonian model is used within Monte-Carlo~\cite{Metropolis} (MC) and molecular dynamics~\cite{PZTmd} simulations to determine the equilibrium microscopic states and dynamics of local electric dipoles in each perovskite five-atom cell of these supercells. The validity of this approach was demonstrated by previous theoretical studies of ultrathin PZT films under compressive strains that (1) yield the vortex stripe domains that periodically alternate along [100] (or along [010])~\cite{Igor,Lai}, in agreement with experimental observation~\cite{obs}; (2) predict a linear dependency between the width of these periodic stripes and the square root of the film's thickness~\cite{Thickness}, as consistent with measurements~\cite{scaling}; and (3) have also led to the prediction of various topological defects~ such as vortices \cite{NaumovLB}, dipolar waves \cite{Sichuga}, bubbles \cite{Lai} and merons (or convex disclinations)~\cite{DrJia} in ferroelectrics, that have been experimentally confirmed \cite{DrJia,Zhang2017,Yadav2016}.

The results presented in Fig.~2 of the manuscript are obtained from Monte Carlo annealing simulations. For each of the considered values of $\beta$ the system is first cooled from 2000~K down to 300~K with 50~K steps under zero external electric field. Then an external electric field with progressively increasing   magnitude is applied. Each run associated with constant values of $\beta$, $T$
and $E$ consists of 40,000 Monte Carlo sweeps with 20,000 sweeps to be considered as thermalization period. For both the cooling and the external electric field simulations the starting configuration for the subsequent parameter value is taken to be the final microscopic state obtained from the preceding run. For molecular dynamics simulations we use a predictor-corrector numerical integration scheme~\cite{PZTmd} with a discrete time step of 0.5 fs.

\begin{figure}[ht!]
\centering
\includegraphics[scale=0.25]{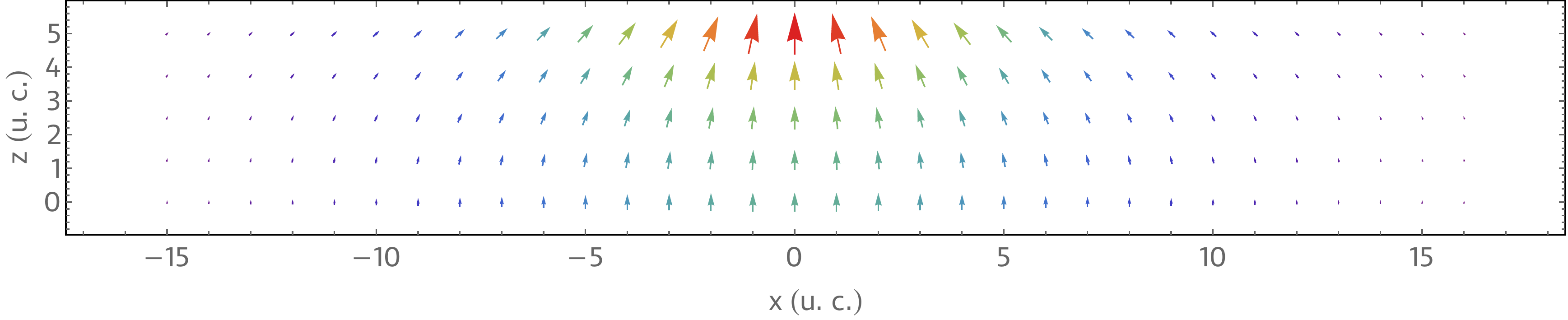}
\caption{\textbf{Driving electric field profile.} Arrows indicate the direction of the position dependent driving electric field within the $y=0$ plane of the $32\times 32\times 5$ supercell for the PFM tip located at $x=y=0$ and $z=5.5$ u.c. Colours from purple to red indicate the increasing field magnitude normalized to its maximum value.
}
\label{fig:pfmfield}
\end{figure}

To demonstrate the gradient driven motion and jump-like displacement of polar skyrmions presented in the manuscript we have used the perturbing potential approximating an electric field generated by a PFM probe. The specific external field model employed here corresponds to the following electric potential~\cite{PFMefield}
\begin{equation}
\begin{split}
\phi=&-\frac{2Q}{4\pi\epsilon_0}\left[\frac{1}{1+\epsilon_1/\epsilon_0}\sum_{n=0}^\infty \frac{\zeta^n}{\sqrt{(z+2nh)^2+r^2}}\right.\\
&-\frac{1-\epsilon_1/\epsilon_2}{(1+\epsilon_1/\epsilon_0)(1+\epsilon_1/\epsilon_2)}\\
&\times\left.\sum_{n=0}^\infty\frac{\zeta^n}{\sqrt{(2(n+1)h-z)^2+r^2}}\right],\\
\zeta=&\frac{(1-\epsilon_1/\epsilon_0)(1-\epsilon_1/\epsilon_2)}{(1-\epsilon_1/\epsilon_0)(1-\epsilon_1/\epsilon_2)},
\end{split}
\end{equation}
where $h$ denotes the film thickness, $\epsilon_0$,$\epsilon_1$ and $\epsilon_2$ denote the vacuum, film and the substrate dielectric permittivities, respectively. This equation is derived under the assumption of the contact mode operation of the tip (the tip located at the surface of the film). The constant $Q$ is determined so as to assure the proper normalization of the corresponding electric field magnitude distribution. Specifically, we require the maximum value of $|\pmb{\nabla}{\phi}|$ to be equal to the specified magnitude $E$ of the perturbing electric field. An example of the resulting distribution of local driving electric field is shown in Fig.~\ref{fig:pfmfield}. Note that the total external electric field at each lattice site corresponds to the sum of the bias background field constant within the supercell volume and the position-dependent electric field perturbation. 

\begin{figure}[ht!]
\centering
\includegraphics[scale=0.5]{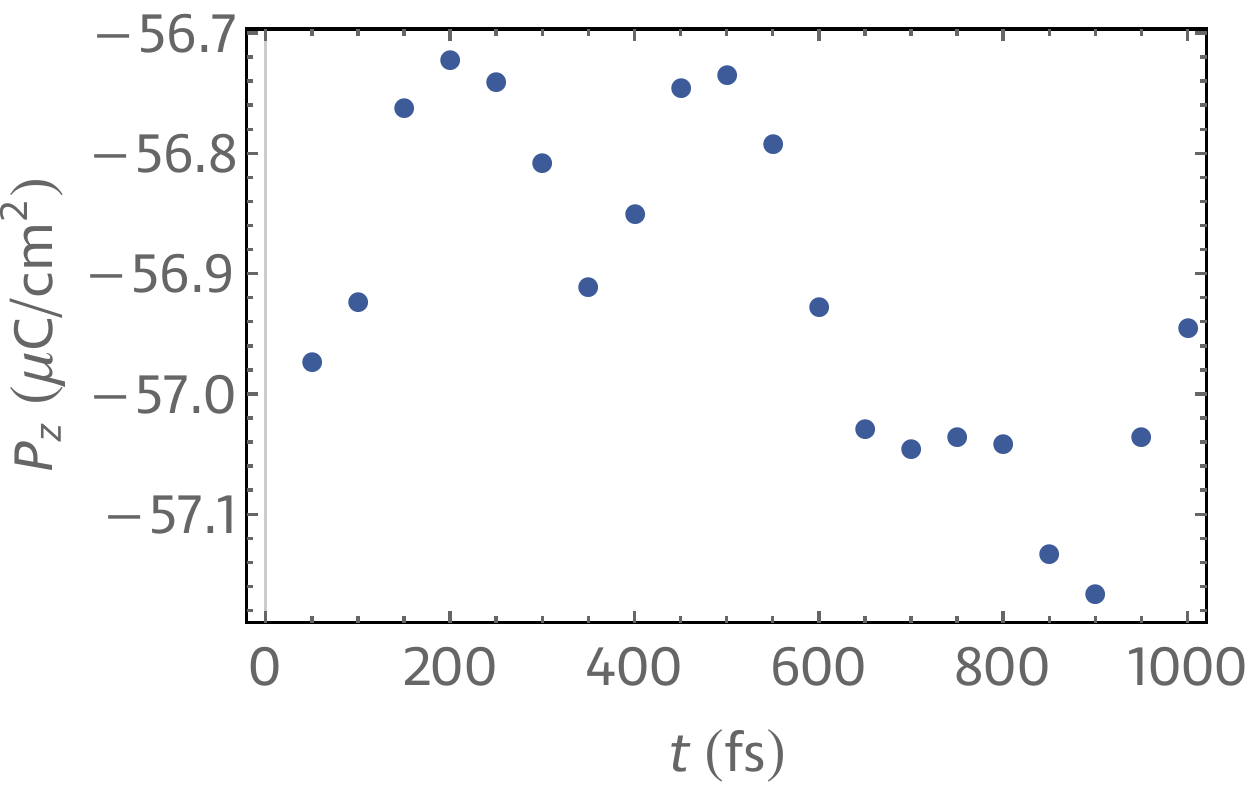}
\caption{\textbf{Conservation of polarization during the skyrmion jump.} Time evolution of the $z$ component of polarization as obtained from molecular dynamics simulations during the skyrmion jump process (see panel (b) of Fig.\ref{fig:motion} and the corresponding discussion). As it can be seen, despite the coexistence of two bubble domains during the jump, the variation of polarization from its average value does not exceed 2\%.
}
\label{fig:pzwithtime}
\end{figure}

We also introduce a correction of the background field in order to reduce the artefacts related to the finite supercell size under in-plane periodic boundary conditions. Specifically, within a macroscopic system, the average electric field generated by a PFM probe is negligibly small as a result of the localized character of the corresponding external potential. Hence, one expects no significant coupling of such electric field perturbation to the overall polarization. However, in a finite supercell such coupling can be strong enough to alter the polarization and the state of system. In order to reduce this artefact we reduce the constant bias by the supercell average of the PFM-tip induced electric field. Note that such correction does not lead to significant alternation of the driving electric field lines (Fig.~\ref{fig:pfmfield}). As a result, the average value of the $z-$component of polarization does not exhibit systematic drift with time as shown in Fig.~\ref{fig:pzwithtime}.

\newpage

\begin{thebibliography}{9}
\bibitem{Rossler2006} R\"{o}{\ss}ler, U. K., Bogdanov, A. N. \& Pfleiderer, C. Spontaneous skyrmion ground states in magnetic metals. Nature 442, 797 (2006).
\bibitem{Igor} Kornev, I.,  Fu, H., \& Bellaiche, L. Ultrathin Films of Ferroelectric Solid Solutions under a Residual Depolarizing Field. \textit{Phys. Rev. Lett.} \textbf{93}, 196104 (2004)
\bibitem{Lai} Lai, B.-K. \textit{et al.}
Electric-Field-Induced Domain evolution in ferroelectric ultrathin films. \textit{Phys. Rev. Lett.} \textbf{96}, 137602 (2006).
\bibitem{Zhang2017} Zhang, Q. \textit{et al.} Nanoscale bubble domains and topological transitions in ultrathin ferroelectric films. \textit{Adv. Mater.} \textbf{29}, 1702375 (2017).
\bibitem{Ramesh2019} Das, S. \textit{et al.} 
\textit{Observation of room-temperature polar skyrmions}. Nature \textbf{568}, 368 (2019).
\bibitem{Zhang2019} Zhang, Q. \textit{et al.} Deterministic Switching of Ferroelectric Bubble Nanodomains. \textit{Adv. Func. Mater.} 1808573 (2019)
\bibitem{racetrack} Parkin, S. S. P., Hayashi, M. \& Thomas, L. Magnetic domain-wall racetrack memory. \textit{Science} \textbf{320}, 190 (2008).
\bibitem{algTopology} Renteln, P. \textit{Manifolds, Tensors, and Forms: An Introduction for Mathematicians and Physicists}. (p. 253, Cambridge Univ. Press.2013)
\bibitem{Yousra2015} Nahas, Y. \textit{et al.} Discovery of stable skyrmionic state in ferroelectric nanocomposites. Nat. Commun. \textbf{6}, 8542 (2015).
\bibitem{DepField} Ponomareva, I.,  Naumov,  I. I., Kornev, I.,  Fu, H. \&  Bellaiche, L. Atomistic treatment of depolarizing energy and field in ferroelectric nanostructures. Phys. Rev. B 72, 140102(R) (2005)
\bibitem{SkyrmionsBook}  Liu, J. P.,  Zhang, Zh.,  Zhao, G. Skyrmions: Topological Structures, Properties, and Applications. CRC Press, Boka Raton (2017).
\bibitem{boojums1} Trickey, S., Adams, E. \&  Dufty, J. \textit{Quantum Fluids and Solids} (Plenum Press, New York, 1977)
\bibitem{boojums3}  Volovik, G. \&  Lavrentovich, O. \textit{Topological dynamics of defects: boojums in nematic drops} \textit{Zh. Eksp. Teor. Fiz.} \textbf{85}, 1997 (1983)
\bibitem{LandauED}  Landau, L. D.,  Pitaevskii, L. P.,  Lifshitz, E.M. Electrodynamics of Continuous Media,  Butterworth-Heinemann ,Oxford (1984).
\bibitem{Yadav2016} Yadav, A. K. \textit{et al.} Observation of polar vortices in oxide superlattices. \textit{Nature}, \textbf{530}, 198 (2016)
\bibitem{PFMefield} Wang, B. \& Woo, C. H. Atomic force microscopy-induced electric field in ferroelectric thin films. \textit{J. Appl. Phys.} \textbf{94}, 4053 (2003)
\bibitem{BinAFE} Xu, B., Iniguez, J. \& Bellaiche, L. Designing lead-free antiferroelectrics for energy storage. \textit{Nat. Commun.} \textbf{8}, 15682 (2017)
\bibitem{Kalinin}  Kalinin, S. V., Jesse, St., Tselev, A., Baddorf A. P. \& Balke, N. The Role of Electrochemical Phenomena in Scanning Probe Microscopy of Ferroelectric Thin Films. \textit{ACS Nano} \textbf{57}, 5683 (2011)


\bibitem{LB1} Bellaiche, L., Garcia, A. 2 $\&$ Vanderbilt, D. Finite-Temperature Properties of Pb(Zr$_{1-x}$Ti$_x$)O$_3$ Alloys from First Principles. \textit{Phys. Rev. Lett.} \textbf{84}, 5427 (2000).
\bibitem{Metropolis} Metropolis, N., Rosenbluth, A. W., Rosenbluth, M. N. \& Teller, A. H. Equation of state calculations by fast computing machines. J. Chem. Phys. 21, 1087 (1953).
\bibitem{PZTmd}   Rapaport, D. \textit{The Art of Molecular Dynamics Simulation.}  (Cambridge University Press, 2001)
\bibitem{obs} Streiffer, S. K., Eastman, J. A., Fong, D. D., Thompson, C., Munkholm, A., Ramana Murty, M. V., Auciello, O., Bai, G. R.  $\&$ Stephenson, G. B. Observation of nanoscale 180 degrees stripe domains in ferroelectric PbTiO$_3$ thin films. \textit{Phys. Rev. Lett.} \textbf{89}, 067601 (2002).

\bibitem{Thickness} Lai, B.-K., Ponomareva, I., Kornev, I. A., Bellaiche, L. $\&$ Salamo, G. J. Thickness dependency of 180 degree stripe domains in ferroelectric ultrathin films: a first-principles study. \textit{Appl. Phys. Lett.} \textbf{91}, 152909 (2007).

\bibitem{scaling} Schilling, A., Adams, T. B., Bowman, R. M., Gregg, J. M.,  Catalan, G. $\&$  Scott, J. F. Scaling of domain periodicity with thickness measured in BaTiO$_3$ single crystal lamellae and comparison with other ferroics. \textit{Phys. Rev. B} \textbf{74}, 024115 (2006).

\bibitem{NaumovLB} Naumov, I. I., Bellaiche, L. $\&$ Fu, H. Unusual phase transitions in ferroelectric nanodisks and nanorods. \textit{Nature} \textbf{432}, 737 (2004).


\bibitem{Sichuga} Sichuga, D., $\&$ Bellaiche, L.  Epitaxial Pb(Zr,Ti)O$_3$ Ultrathin Films under Open-Circuit Electrical Boundary conditions. \textit{Phys. Rev. Lett.} \textbf{106},  196102 (2011).



\bibitem{DrJia} Lu, L., Nahas, Y., Liu, M., Du, H., Jiang, Z., Ren, S., Wang, D., Jin, L., Prokhorenko, S., Jia C.-L. $\&$ Bellaiche, L.  Topological defects with distinct dipole configurations in PbTiO$_3$-SrTiO$_3$ multilayer films. \textit{Phys. Rev. Lett.} \textbf{120}, 177601 (2018).
%
%


%

\end{thebibliography}
\end{document}